\documentclass[11pt]{article}

\usepackage[final]{acl}

\usepackage{times}
\usepackage{latexsym}
\usepackage{amsmath} 
\usepackage{amssymb} 
 \usepackage{booktabs} 
 \usepackage{booktabs}
\usepackage{multirow}
\usepackage{makecell}
\usepackage{colortbl}
\usepackage[table]{xcolor}
\usepackage{makecell}
\usepackage{pgfplots}
\pgfplotsset{compat=1.18}
\definecolor{oursrow}{RGB}{230,241,251}   % light blue — our model rows
\definecolor{bestval}{RGB}{12,68,124}     % dark blue — bold best numbers
\newcommand{\best}[1]{\textcolor{bestval}{\textbf{#1}}}
\newcommand{\ours}{\rowcolor{oursrow}}
\newcommand{\ph}{--}   % placeholder for numbers not yet known

\usepackage{tcolorbox}
\tcbuselibrary{listings,breakable}

\definecolor{promptheader}{RGB}{139,69,19}
\definecolor{promptbg}{RGB}{253,248,244}
\definecolor{jsonbg}{RGB}{245,245,245}

\newtcolorbox{promptbox}[1]{
  title=#1,
  fonttitle=\small\bfseries\color{white},
  colbacktitle=promptheader,
  colback=promptbg,
  coltitle=white,
  boxrule=0.4pt,
  arc=4pt,
  breakable,
  left=8pt, right=8pt, top=6pt, bottom=6pt
}
\usepackage[T1]{fontenc}
\usepackage[utf8]{inputenc}

\usepackage{microtype}

\usepackage{inconsolata}

\usepackage{graphicx}

\usepackage[table]{xcolor}
\definecolor{myasd}{RGB}{225,236,248}
\title{MEMORA: A Memory-Enhanced Multimodal Committee Reranking Agent for Reasoning-Intensive Retrieval}

\author{
  \textbf{Mohamed Mahmoud\textsuperscript{1}, Mostafa Farouk Senussi\textsuperscript{1}, Mahmoud Abdalla\textsuperscript{1},} \\
  \textbf{Mahmoud SalahEldin Kasem\textsuperscript{1}, Abdelrahman Abdallah\textsuperscript{2}, Hyun Soo Kang\textsuperscript{1}} \\
  \textsuperscript{1}Chungbuk National University \quad \textsuperscript{2}University of Innsbruck \\
  \texttt{hskang@cbnu.ac.kr}
}

\begin{document}
\maketitle

\begin{abstract}
Reasoning-intensive retrieval is fundamentally challenged when queries 
demand deep semantic inference and arrive in multimodal form. Existing 
systems treat each query in isolation, rely on monolithic rerankers, and 
lack retrievers trained for cross-modal reasoning. We introduce 
\textbf{MEMORA}\footnote{Code and data are available: \url{https://github.com/HarnessLab/MEMORA}.}, a \textbf{M}emory-\textbf{E}nhanced \textbf{M}ultimodal 
\textbf{O}rchestrated \textbf{R}eranking \textbf{A}gent built around 
three contributions: (1)~\textbf{Collaborative Factor-Committee Reranking}, 
which decomposes queries into intent factors, extracts factor-specific 
evidence into a structured matrix, and synthesizes judgments via a 
memory-conditioned meta-verdict model; (2)~an \textbf{Episodic Memory 
Module} that tracks factor outcomes across queries and dynamically adjusts 
priorities at inference time; and (3)~a \textbf{Reasoning-Aware Dense 
Retriever} fine-tuned with chain-of-thought augmentation on 20k$+$ 
reasoning-intensive pairs. Evaluated on MM-BRIGHT across 29 domains, 
\textsc{MEMORA} achieves nDCG@10 of \textbf{41.5} (text-to-text) and 
\textbf{38.9} (multimodal query-to-text), outperforming the strongest 
agentic text-only baseline by \textbf{+8.3} points and the strongest 
agentic multimodal baseline by \textbf{+7.9} points.
\end{abstract}

\section{Introduction}

Retrieval systems are increasingly expected to handle queries that demand 
genuine reasoning rather than surface-level matching. This challenge is 
compounded in real-world settings, where users combine natural language 
with images --- attaching circuit diagrams to technical questions, 
screenshots of error logs to debugging requests, or scientific figures 
to research queries. In these settings, identifying the relevant document 
requires understanding what the image depicts, how it relates to the 
textual question, and which documents in the corpus collectively address 
both dimensions. Yet most retrieval systems are designed for text-only 
queries, and those that support multimodal input lack the reasoning 
capacity to handle this level of cross-modal inference.

This gap is starkly exposed by MM-BRIGHT~\cite{abdallah2026mm}, the first 
benchmark for reasoning-intensive multimodal retrieval. Despite significant 
advances in vision-language models, the best multimodal encoder achieves 
only 27.6 nDCG@10 on the multimodal-to-text task --- \textit{lower} than 
the best text-only retriever (32.2), meaning that adding visual information 
actively \textit{hurts} performance. This counterintuitive result reveals 
three compounding failure modes that existing approaches have not addressed 
jointly: 
(a)~\textbf{Monolithic reranking}: existing rerankers apply a single, 
uniform reasoning pass over all candidates, regardless of how heterogeneous 
the query's underlying information needs are; 
(b)~\textbf{Stateless retrieval}: systems treat every query independently, 
discarding accumulated knowledge about which retrieval strategies succeed 
or fail within a domain; and 
(c)~\textbf{Weak multimodal grounding}: standard dense encoders are not 
trained to handle the abstract, cross-modal reasoning that multimodal 
queries demand, producing embeddings where visual signals add noise rather 
than signal. Prior work addresses at most one of these failures in 
isolation~\cite{sun2023chatgpt,weller2025rank1,wang2023query2doc,
gao2023precise,zhang2024gme,jiang2024vlm2vec}.

\begin{figure*}[t]
    \centering
    \includegraphics[width=\linewidth]{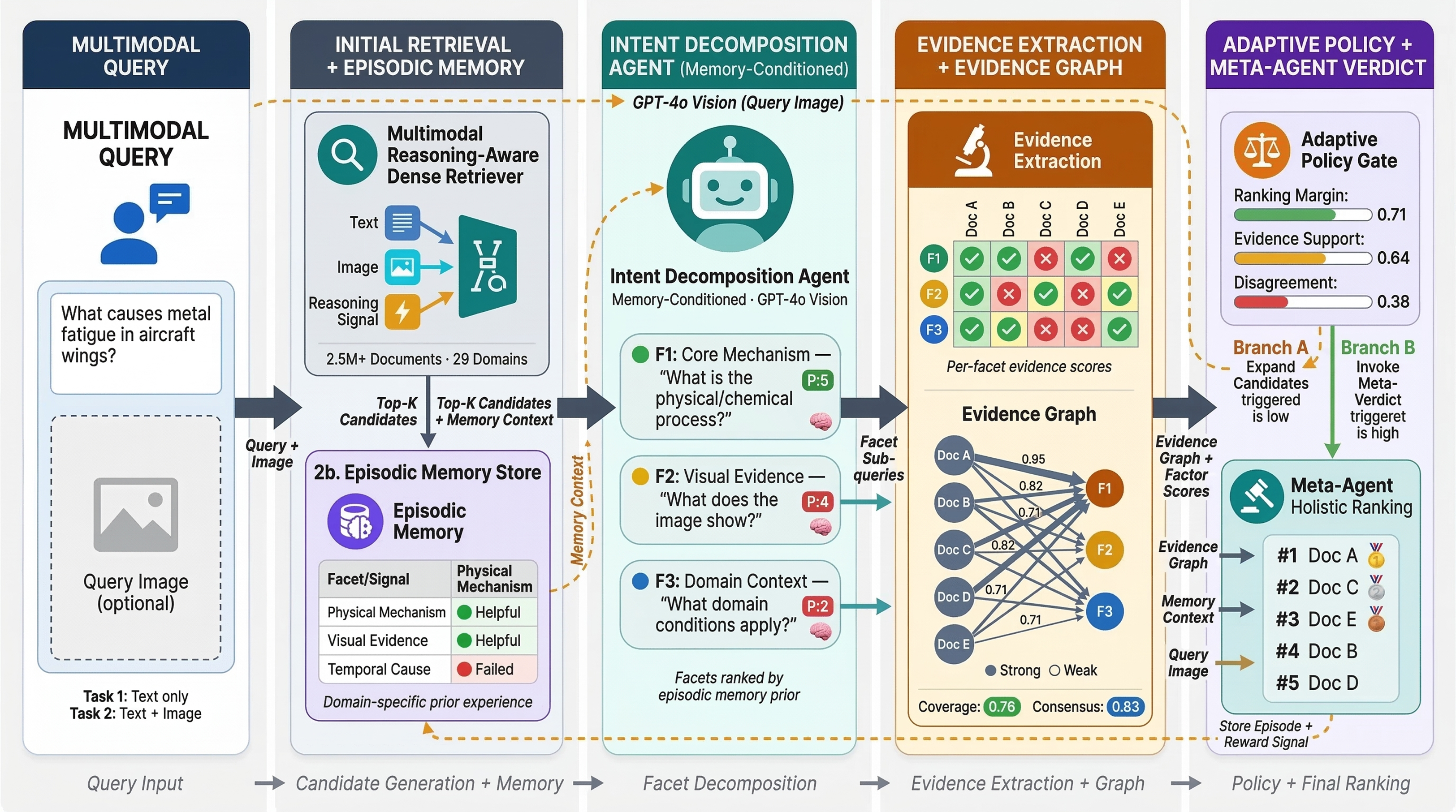}
    \caption{Overview of \textsc{MEMORA}. Given a multimodal query $(t, x)$, 
    the Episodic Memory Module retrieves domain-specific prior experience to 
    guide factor decomposition and candidate expansion. The Multimodal 
    Reasoning-Aware Dense Retriever produces an initial candidate pool, which 
    is expanded via factor-specific subqueries. A factor evidence extractor 
    then builds a bipartite factor-document evidence graph, whose structured 
    signals feed into a memory-conditioned meta-verdict model for final 
    ranking. An Adaptive Controller decides at each step whether to expand 
    retrieval, invoke the judge, or stop.}
    \label{fig:overview}
\end{figure*}

We introduce \textbf{MEMORA} (\textbf{M}emory-\textbf{E}nhanced 
\textbf{M}ultimodal \textbf{O}rchestrated \textbf{R}eranking 
\textbf{A}gent), a unified agentic framework that addresses all three 
failure modes jointly. As illustrated in Figure~\ref{fig:overview}, 
MEMORA operates through four tightly integrated stages: 
(1)~a \textbf{Reasoning-Aware Dense Retriever} encodes queries enriched 
with chain-of-thought traces to produce a high-quality initial candidate 
pool; (2)~an \textbf{Episodic Memory Module} retrieves similar past 
queries from domain memory and uses accumulated factor outcome experience 
to decompose the current query into prioritized intent factors; 
(3)~a \textbf{Collaborative Factor Committee} issues factor-specific 
subqueries, extracts targeted evidence per candidate, and organizes 
judgments into a bipartite \textbf{factor-document evidence matrix} that 
a memory-conditioned meta-verdict model consults to produce the final 
ranking; and (4)~an \textbf{Adaptive Controller} dynamically decides 
whether to expand retrieval, invoke the judge, or stop based on ranking 
margin and evidence support, making MEMORA responsive to query difficulty 
rather than fixed in its computation.

Our contributions are as follows:
\begin{itemize}
    \item We identify three compounding failure modes in reasoning-intensive 
    multimodal retrieval --- monolithic reranking, stateless retrieval, and 
    weak multimodal grounding --- and propose \textbf{MEMORA}, a unified 
    agentic framework that addresses all three jointly (\S\ref{sec:method}).

    \item We introduce \textbf{Collaborative Factor-Committee Reranking}, 
    in which query intent is decomposed into factors, evidence is extracted 
    per factor and organized into a bipartite factor-document graph, and a 
    memory-conditioned meta-verdict model produces the final ranking 
    (\S\ref{sec:method}).

    \item We propose an \textbf{Episodic Memory Module} that accumulates 
    domain-specific factor outcome experience across queries, dynamically 
    adjusting factor priorities at inference time without retraining, and 
    train a \textbf{Reasoning-Aware Dense Retriever} on 20k$+$ 
    reasoning-intensive Stack Exchange pairs with CoT annotation and hard 
    negatives (\S\ref{sec:retriever}).

    \item The full \textsc{MEMORA} system achieves \textbf{41.5} and
    \textbf{38.9} nDCG@10 on MM-BRIGHT Tasks~1 and~2, outperforming
    the strongest agentic text-only baseline by \textbf{+8.3} points
    and the strongest agentic multimodal baseline by \textbf{+7.9}
    points (\S\ref{sec:retriever_results}).
\end{itemize}

\section{Related Work}
\label{sec:related_work}

\paragraph{Retrieval benchmarks and models.}
Early benchmarks such as BEIR~\cite{thakur2021beir} and
MTEB~\cite{muennighoff2023mteb} primarily measure surface-level relevance.
BRIGHT~\cite{su2024bright} and RAR-b~\cite{xiao2024rar} made reasoning a
core requirement and exposed sharp performance drops in existing systems.
MM-BRIGHT~\cite{abdallah2026mm} extends this setting to multimodal queries,
and the concurrent benchmarks MRMR~\cite{zhang2025mrmr} and
ARK~\cite{lin2026ark} confirm that visual and spatial reasoning remain
persistent bottlenecks. Dense retrievers such as
E5~\cite{wang2024improving}, GritLM~\cite{muennighoff2024generative}, and
SFR-Embedding~\cite{meng2024sfrembedding} perform well on standard
benchmarks but degrade when relevance depends on reasoning.
ReasonIR~\cite{shao2025reasonir} and DiVeR~\cite{long2025diver} partially
close this gap through reasoning-aware training. Multimodal retrievers,
including VLM2Vec~\cite{jiang2024vlm2vec}, GME~\cite{zhang2024gme},
Argus~\cite{abdallah2026argus}, and Nomic-Vision~\cite{nussbaum2024nomic},
extend dense retrieval to vision-language inputs. However, their best
result on MM-BRIGHT Task~2 (27.6 nDCG@10) still falls below the strongest
text-only baseline, indicating that visual signals introduce noise unless
they are accompanied by explicit reasoning.

\paragraph{Multimodal reasoning-intensive retrieval pipelines.}
A recent line of work addresses this gap by inserting explicit reasoning
between the multimodal query and the retriever. BRIDGE~\cite{mounis2026bridge}
reformulates multimodal queries into text queries aligned with text
retrievers, learning this alignment through reinforcement learning.
HIVE~\cite{abdalla2026hive} adopts a query--hypothesize--verify loop in
which an LLM proposes candidate hypotheses and checks them against
retrieved evidence. MARVEL~\cite{kasem2026marvel} combines adaptive query
expansion with reranking in a unified expand--rerank--retrieve pipeline,
while VISA-Agent~\cite{abdalla2026visa} converts visual content into
symbolic representations that an agent can reason over during retrieval.
Together, these methods show that translating visual content into
reasoning-friendly forms is more effective than embedding it directly.

\paragraph{Reranking, query expansion, and memory.}
LLM-based rerankers~\cite{abdallah2025dear,zhuang2023rankt5,abdallah2026bracketrank,pradeep2023rankzephyr}
substantially outperform dense retrieval alone. RankGPT~\cite{sun2023chatgpt}
pioneered listwise reranking, and Rank1~\cite{weller2025rank1} and
RankR1~\cite{zhuang2025rank} further incorporate chain-of-thought reasoning
and reinforcement learning. Query expansion
methods~\cite{gao2023precise,wang2023query2doc,lei2025thinkqe} and agentic
systems~\cite{trivedi2023interleaving,chen2026agentir} improve recall
through LLM-generated subqueries and interleaved reasoning.
Memory-augmented approaches draw on episodic experience for
LLMs~\cite{fountas2024human}, case-based RAG~\cite{wiratunga2024cbr}, and
multimodal episode indexing~\cite{xie2024embodied}. Despite this progress,
no prior work addresses all three failure modes jointly: existing rerankers
are monolithic, retrievers are stateless, and no system unifies episodic
factor memory with committee-based reranking under multimodal conditions.

\section{Method}
\label{sec:method}

\textsc{MEMORA} is a memory-conditioned retrieval agent that treats each 
query $q=(t,x)$ --- text $t$ and optional image $x$ --- as a composition 
of latent intent factors. Given a corpus $\mathcal{D}=\{d_j\}_{j=1}^N$, 
the goal is to produce a ranking $\pi(q)=(d_{(1)},\dots,d_{(K)})$ that 
maximizes relevance to the full user intent. Unlike fixed prompt pipelines, 
\textsc{MEMORA} persistently stores which factors helped or failed in 
previous queries and uses this signal to guide both retrieval and ranking.

\subsection{Overview}
The pipeline proceeds in four stages: (1)~a reasoning-aware dense retriever 
produces an initial candidate pool; (2)~domain memory episodes guide 
decomposition of the query into intent factors; (3)~factor-specific subqueries 
expand the pool and a factor-document evidence graph is constructed; 
(4)~a factor-aware scorer ranks candidates, with a memory-aware listwise 
judge invoked when needed.

\subsection{Reasoning-Aware Dense Retriever}
\label{sec:retriever}

Standard dense retrievers are not trained for the deep semantic inference 
that reasoning-intensive queries demand. We therefore train a dedicated 
retriever on a purpose-built corpus of reasoning-intensive pairs as the 
backbone of \textsc{MEMORA}.

\paragraph{Architecture.}
We adopt a bi-encoder built on \textbf{Qwen-Embedding}. The query encoder 
$\phi_q$ produces $\mathbf{q} = \phi_q(q) \in \mathbb{R}^d$; document 
embeddings $\mathbf{d}_j = \phi_d(d_j)$ are pre-computed offline. 
Retrieval scores are:
\begin{equation}
s_j^{(0)} = \mathbf{q}^\top \mathbf{d}_j.
\end{equation}

\paragraph{Training Data.}
We construct \textbf{20{,}000$+$ reasoning-intensive question--answer pairs} 
from Stack Exchange, where each question is augmented with a GPT-4o 
chain-of-thought annotation articulating the latent retrieval intent. 
Hard negatives are mined via an initial BM25 pass.

\paragraph{Reasoning Signal.}
The query encoder receives the concatenation $[r;\,q]$, where $r$ is the 
CoT annotation, teaching it to embed latent retrieval intent rather than 
surface features. At inference time the encoder operates on the raw query 
alone, having internalized this behavior during training.

\paragraph{Training.}
We fine-tune with in-batch and hard negatives via InfoNCE:
\begin{equation}
\mathcal{L} = -\log \frac{
    \exp\!\left(\mathbf{q}^\top \mathbf{d}^+ / \tau\right)
}{
    \exp\!\left(\mathbf{q}^\top \mathbf{d}^+ / \tau\right)
    + \sum_{j} \exp\!\left(\mathbf{q}^\top \mathbf{d}^-_j / \tau\right)
},
\end{equation}
where $\mathbf{d}^+$ is the positive document, $\mathbf{d}^-_j$ are hard 
negatives, and $\tau$ is a temperature hyperparameter.

\subsection{Initial Retrieval}
The top-$K_0$ documents scored by $R$ form the initial candidate set 
$\mathcal{C}^{(0)}$. For multimodal queries, $R$ consumes both $t$ and 
$x$; for text-only queries, $x$ is omitted.

\subsection{Domain Memory}

For each domain $c$, \textsc{MEMORA} maintains an episodic memory 
$\mathcal{M}_c = \{e_i\}_{i=1}^{T_c}$. Each memory entry stores the previous 
query, its factor graph, evidence snippets, factor scores, factor outcomes, 
and final reward:
\begin{equation}
e_i = \big(q_i,\, \mathcal{G}_i,\, \mathcal{S}_i,\, \Delta_i,\, r_i\big).
\end{equation}
Here, $\mathcal{G}_i$ is the stored factor graph, $\mathcal{S}_i$ contains 
factor scores per document, and $\Delta_i$ summarizes which factors helped 
or failed.

Given a new query $q$, we retrieve the most similar episodes from the same 
domain using a lightweight hybrid similarity:
\begin{equation}
\operatorname{sim}(q, e_i)
=
\lambda_J\, J\!\big(T(q), T(q_i)\big)
+
(1 - \lambda_J)\, \rho(q, q_i),
\end{equation}
where $T(\cdot)$ extracts content tokens, $J(\cdot,\cdot)$ is Jaccard 
similarity, and $\rho(\cdot,\cdot)$ is a string-level similarity score. 
The top-$K_m$ episodes form the retrieved memory set 
$\mathcal{M}(q) \subseteq \mathcal{M}_c$.

From $\mathcal{M}(q)$, we compute two types of priors. The first is a 
document prior
\begin{equation}
m_j =
\frac{1}{|\mathcal{M}(q)|}
\sum_{e_i \in \mathcal{M}(q)}
\mathbb{I}[d_j \in \operatorname{TopDocs}(e_i)],
\end{equation}
which tracks documents that repeatedly appeared in related episodes;
this prior is only beneficial when combined with factor outcome signals,
as it can otherwise introduce frequency bias toward common documents
over genuinely relevant ones (see \S\ref{sec:memory_analysis}).
The second is a factor prior: for each factor label $f_k$, memory
returns a helpfulness prior $h_k$ and a failure prior $\ell_k$,
estimated from previous outcomes stored in $\Delta_i$.

\subsection{Memory-Conditioned Factor Decomposition}

Instead of treating the query as a single unit, \textsc{MEMORA} decomposes 
it into a small set of intent factors
\begin{equation}
F(q) = \{f_1, \dots, f_K\}.
\end{equation}
Each factor $f_k$ contains a factor name, a focused factor question, a 
retrieval-oriented subquery, and a priority score. The number of factors $K$ 
is itself adaptive: it is set by the controller based on the number of memory 
hits $|\mathcal{M}(q)|$ and whether an image is present, within the range 
$[K_{\min}, K_{\max}]$. The factor planner is conditioned on the query and 
the retrieved memory summary, so the decomposition is informed by both the 
current input and past domain experience. For multimodal queries, the factor 
planner is additionally conditioned on a visual description of $x$ produced 
by a vision-language model (GPT-4o), enabling factors to explicitly reference 
visual content.

Let $p_k \in \{1,\dots,5\}$ denote the planner's initial priority for factor 
$f_k$. We then adjust this priority using memory:
\begin{equation}
\tilde{p}_k
=
\operatorname{clip}
\!\Big(
p_k
+ \mathbb{I}[h_k > \tau_+]
- \mathbb{I}[\ell_k > \tau_-],
\; 1,\; 5
\Big).
\end{equation}
This simple rule increases the weight of factors that have historically helped 
and suppresses factors that repeatedly over-supported distractors.

\subsection{Factor-Guided Retrieval Expansion}

The initial pool $\mathcal{C}^{(0)}$ may miss documents needed by one 
specific factor. To improve recall, \textsc{MEMORA} may issue factor-specific 
retrieval calls. For factor $f_k$, let $z_k$ denote its subquery. The 
retriever returns
\begin{equation}
s_{j,k}^{\mathrm{exp}} = R(z_k, d_j).
\end{equation}
We normalize all retrieval scores and combine the initial retrieval, factor 
expansion, and memory prior into a single candidate score:
\begin{equation}
\hat{s}_j
=
\max\!\Big(
\bar{s}_j^{(0)},\;
\max_{k}\, \bar{s}_{j,k}^{\mathrm{exp}}
\Big)
+ \lambda_m\, m_j.
\end{equation}
The refined candidate set is then
\begin{equation}
\mathcal{C}
=
\operatorname{TopK}_{K_c}\!\big(\{\hat{s}_j\}_{j=1}^N\big).
\end{equation}

\subsection{Factor Evidence Extraction}

For each candidate document $d_j \in \mathcal{C}$ and each factor $f_k$, 
an evidence extractor returns a factor relevance score 
$a_{j,k} \in [0, 10]$ and a short evidence snippet $u_{j,k}$:
\begin{equation}
(a_{j,k},\, u_{j,k}) = E(q,\, f_k,\, d_j).
\end{equation}
Scores are normalized within each factor:
\begin{equation}
\bar{a}_{j,k}
=
\frac{a_{j,k}}{\max_{j'} a_{j',k} + \epsilon}.
\end{equation}
This step yields factor-specific evidence rather than a single opaque 
relevance judgment.

\subsection{Factor Evidence Graph}

The core representation in \textsc{MEMORA} is a bipartite graph
$\mathcal{G} = (\mathcal{V}_d,\, \mathcal{V}_f,\, \mathcal{E})$,
where document nodes $\mathcal{V}_d = \{d_j\}$ connect to factor nodes 
$\mathcal{V}_f = \{f_k\}$. An edge $(d_j, f_k) \in \mathcal{E}$ carries two attributes: the 
normalized factor score $\bar{a}_{j,k}$ and the extracted snippet 
$u_{j,k}$. The edge weight $w_{j,k} = \bar{a}_{j,k}$ encodes how 
strongly document $d_j$ satisfies factor $f_k$, enabling the 
meta-verdict judge to consult structured per-factor support rather 
than a single opaque relevance score.

\subsection{Factor-Aware Document Scoring}

A good document should satisfy multiple factors simultaneously. We therefore 
combine four signals: average factor support, factor coverage, factor 
consensus, and retrieval prior. For document $d_j$, we define
\begin{equation}
u_j = \frac{1}{K}\sum_{k=1}^{K} \bar{a}_{j,k},
\end{equation}
\begin{equation}
c_j = \frac{1}{K}\sum_{k=1}^{K} \mathbb{I}[\bar{a}_{j,k} \ge \tau],
\end{equation}
\begin{equation}
v_j = \min_{k}\, \bar{a}_{j,k},
\end{equation}
where $u_j$ measures overall support, $c_j$ measures how many factors are 
covered, and $v_j$ penalizes documents that fail any single critical factor. 
The final factor-based score is
\begin{equation}
s_j
=
\lambda_u\, u_j
+ \lambda_c\, c_j
+ \lambda_v\, v_j
+ \lambda_r\, \hat{s}_j,
\end{equation}
where the weights $\lambda_u, \lambda_c, \lambda_v, \lambda_r$ balance 
structural evidence against retrieval confidence.

\subsection{Adaptive Controller}

\textsc{MEMORA} does not execute the same action sequence for every query. 
Instead, an adaptive controller decides whether to expand retrieval, invoke 
the final judge, or stop. At step $t$, the controller observes a compact 
state vector $z_t \in \mathbb{R}^5$: normalized ranking margin, memory hit 
ratio, mean evidence support, Spearman correlation between retriever-based 
and factor-based rankings, and a binary image indicator. The action is 
selected as
\begin{equation}
a_t = \arg\max_{a \in \mathcal{A}}\, g_a(z_t), \quad
g_a(z_t) = w_a^\top z_t + b_a,
\end{equation}
where $\mathcal{A} = \{\texttt{expand}, \texttt{judge}, \texttt{stop}\}$.
The weights $w_a$ and biases $b_a$ are set by cross-validation on a 
held-out development split. The linear form is intentional: it is 
interpretable, adds negligible inference overhead, and can be replaced 
with a learned policy in future work.

\begin{table*}[t]
\centering
\caption{Task~1 (text-to-text) nDCG@10 per domain across 29 MM-BRIGHT domains. Each retriever is paired with SelfRAG, Search-o1, and \textsc{MEMORA} as
reranking layers. }
\label{tab:domain_rows_model_cols}
\setlength{\tabcolsep}{3.5pt}
\renewcommand{\arraystretch}{0.82}
\resizebox{\textwidth}{!}{%
\begin{tabular}{l| c c c c|  c c c c| c c c c}
\toprule
\textbf{Domain}
  & \textbf{BM25} & \makecell{\textbf{BM25+}\\\textbf{SelfRAG}}
  & \makecell{\textbf{BM25+}\\\textbf{Search-o1}} & \makecell{\textbf{BM25+}\\\textbf{Memora}}
  & \textbf{E5} & \makecell{\textbf{E5+}\\\textbf{SelfRAG}}
  & \makecell{\textbf{E5+}\\\textbf{Search-o1}} & \makecell{\textbf{E5+}\\\textbf{Memora}}
  & \textbf{DiVeR} & \makecell{\textbf{DiVeR+}\\\textbf{SelfRAG}}
  & \makecell{\textbf{DiVeR+}\\\textbf{Search-o1}} & \makecell{\textbf{DiVeR+}\\\textbf{Memora}} \\
\midrule
\ours \multicolumn{13}{c}{\textit{STEM \& Life Sciences}} \\
\midrule
Acad   &  9.1 & 11.1 & 10.5 & 15.2 & 24.6 & 26.6 & 30.1 & 35.7 & 31.6 & 34.0 & 37.2 & 44.9 \\
Bio    &  4.8 &  6.4 &  7.5 & 10.4 & 18.2 & 21.4 & 27.1 & 36.1 & 27.7 & 32.5 & 40.0 & 54.3 \\
Chem   &  7.8 &  8.8 & 10.2 & 11.8 & 25.4 & 26.6 & 28.8 & 32.1 & 33.8 & 35.2 & 32.8 & 41.8 \\
Phys   &  4.0 &  5.3 &  6.2 & 10.2 & 14.6 & 15.1 & 15.9 & 17.1 & 19.9 & 20.0 & 20.2 & 20.6 \\
Math   &  2.0 &  3.1 &  4.9 & 10.5 & 18.1 & 19.3 & 21.4 & 24.6 & 34.2 & 35.0 & 32.5 & 38.8 \\
Earth  &  4.7 &  5.8 &  7.5 & 11.1 & 29.1 & 31.0 & 34.4 & 39.7 & 34.6 & 36.7 & 40.4 & 46.2 \\
BioAc  &  8.5 & 10.2 & 15.1 & 18.1 & 20.5 & 21.8 & 24.0 & 27.5 & 27.4 & 28.4 & 30.1 & 32.9 \\
BioInf &  5.3 &  6.4 &  8.4 & 10.5 & 18.9 & 19.1 & 19.4 & 19.8 & 32.7 & 32.7 & 31.7 & 29.3 \\
Med    & 12.6 & 13.2 & 11.2 & 15.4 & 34.7 & 36.1 & 38.7 & 42.7 & 36.1 & 37.6 & 40.2 & 44.4 \\
\midrule
\ours \multicolumn{13}{c}{\textit{Software \& Technical Systems}} \\
\midrule
Apple  &  1.0 &  2.0 &  5.1 &  7.2 & 24.8 & 24.9 & 25.1 & 25.3 & 22.8 & 23.0 & 21.3 & 23.8 \\
Ubuntu & 17.4 & 15.1 & 20.0 & 22.0 & 29.0 & 29.8 & 31.2 & 33.3 & 44.6 & 45.3 & 43.5 & 48.5 \\
BTC    &  3.3 &  4.3 &  5.1 &  7.0 & 32.4 & 33.1 & 34.4 & 36.4 & 30.4 & 31.1 & 31.4 & 34.4 \\
Crypto &  0.8 &  3.1 &  6.2 &  8.7 & 10.2 & 10.7 & 11.5 & 12.8 & 24.3 & 24.3 & 21.3 & 18.9 \\
GIS    &  1.3 &  2.1 &  4.5 & 10.5 & 25.0 & 25.0 & 25.0 & 25.0 & 31.7 & 31.7 & 30.7 & 28.4 \\
QC     &  3.8 &  4.2 &  6.2 & 12.5 & 11.4 & 11.7 & 12.4 & 13.3 & 13.0 & 13.1 & 11.2 & 13.5 \\
Robot  &  5.0 &  6.0 &  9.7 & 11.8 & 20.2 & 20.8 & 21.8 & 23.5 & 33.4 & 33.4 & 30.5 & 33.6 \\
Sales  &  3.9 &  4.8 &  3.1 & 12.1 & 25.7 & 27.7 & 31.2 & 36.7 & 38.6 & 40.9 & 41.9 & 51.2 \\
\midrule
\ours \multicolumn{13}{c}{\textit{Social Sciences \& Humanities}} \\
\midrule
Econ   &  4.0 &  5.0 &  6.1 &  9.8 & 27.7 & 29.2 & 31.7 & 35.8 & 31.9 & 33.4 & 31.1 & 40.4 \\
Psych  &  5.3 &  5.4 &  7.9 & 10.7 & 22.8 & 24.4 & 27.3 & 31.8 & 29.3 & 31.2 & 31.5 & 39.7 \\
Phil   &  4.1 &  5.3 &  7.7 & 15.2 & 21.3 & 22.8 & 25.4 & 29.4 & 17.4 & 19.0 & 20.8 & 26.2 \\
Law    &  6.2 &  7.5 & 10.1 & 14.9 & 50.2 & 52.4 & 56.3 & 62.4 & 52.3 & 54.5 & 54.5 & 64.7 \\
Christ & 30.3 & 20.1 & 25.1 & 35.1 & 29.9 & 30.7 & 32.0 & 34.1 & 35.2 & 37.4 & 40.2 & 47.3 \\
Islam  &  9.8 & 10.2 & 12.8 & 17.2 & 27.5 & 28.1 & 29.2 & 31.0 & 34.9 & 35.2 & 31.8 & 36.7 \\
\midrule
\ours \multicolumn{13}{c}{\textit{Applied Domains}} \\
\midrule
Aviat  &  1.1 &  2.2 &  7.1 &  6.8 & 21.9 & 22.6 & 24.0 & 26.1 & 29.5 & 30.1 & 30.3 & 33.1 \\
Game   & 36.0 & 32.1 & 40.2 & 41.0 & 42.1 & 44.5 & 48.9 & 55.7 & 50.4 & 54.9 & 50.0 & 75.6 \\
PM     & 18.0 & 17.0 & 15.1 & 20.7 & 31.0 & 31.0 & 31.0 & 31.0 & 40.4 & 35.4 & 30.4 & 37.8 \\
Sustain& 11.2 & 12.0 & 15.1 & 19.2 & 30.6 & 32.4 & 35.7 & 40.7 & 34.8 & 30.7 & 35.1 & 45.4 \\
Travel & 22.2 & 23.4 & 25.1 & 24.8 & 24.8 & 25.6 & 27.1 & 29.4 & 38.3 & 40.9 & 41.6 & 53.0 \\
Quant  &  2.3 &  5.1 &  7.2 & 10.1 & 22.5 & 24.3 & 27.5 & 32.4 & 23.2 & 25.0 & 28.2 & 33.2 \\
\midrule
\textbf{Avg.} & 8.5 & 8.9 & 11.1 & 14.8 & 25.3 & 26.5 & 28.6 & 31.8
              & 32.2 & 33.2 & 33.2 & \best{39.3} \\
\bottomrule
\end{tabular}%
}
\end{table*}

\subsection{Memory-Aware Final Ranking and Memory Update}

When ambiguity remains after factor scoring, \textsc{MEMORA} invokes a 
listwise meta-verdict judge that receives the query, optional image, top 
candidate documents, the factor evidence graph, and the retrieved memory 
summary. The judge consults the factor graph rather than raw documents 
alone, keeping the final ranking grounded in explicit structured evidence.

After ranking, \textsc{MEMORA} computes which factors were useful and 
writes this signal back to memory. For factor $f_k$, let $\mu_k^+$ and 
$\mu_k^-$ denote the average factor scores over relevant documents 
$\mathcal{D}^+$ and non-relevant candidates $\mathcal{D}^-$:
\begin{equation}
\mu_k^{\pm} =
\frac{1}{|\mathcal{D}^{\pm}|}
\sum_{d_j \in \mathcal{D}^{\pm}} \bar{a}_{j,k}.
\end{equation}
The helpfulness and failure signals are then:
\begin{equation}
\begin{aligned}
\delta_k^+ &= \max(0,\; \mu_k^+ - \mu_k^-), \\
\delta_k^- &= \max(0,\; \mu_k^- - \mu_k^+).
\end{aligned}
\end{equation}
These are written back to domain memory as the helpfulness and failure 
priors $h_k$ and $\ell_k$ for future queries, closing the memory loop 
illustrated in Figure~\ref{fig:overview}. 
% W

\section{Experiments}

\subsection{Dataset}

We evaluate on MM-BRIGHT~\cite{abdallah2026mm}, the first reasoning-intensive
multimodal retrieval benchmark, comprising \textbf{2,803 queries} across
\textbf{29 technical domains} in four categories (STEM \& Life Sciences,
Software \& Technical Systems, Social Sciences \& Humanities, and Applied
Domains) over a corpus of \textbf{2.5M documents} from Stack Exchange.
We evaluate on \textbf{Task~1} (text-to-text) and \textbf{Task~2}
(multimodal query-to-text, where queries include diagrams, charts, and
screenshots), using \textbf{nDCG@10} as the primary metric.

To prevent training-test contamination, we constructed MEMORA-R's
training set by excluding all Stack Exchange threads whose post IDs
appear in the MM-BRIGHT document corpus, as well as any threads
posted after the benchmark's data collection cutoff. The resulting
training split is temporally and ID-disjoint from both the retrieval
corpus and the query set used for evaluation.

\subsection{Experimental Setup}
The retriever was trained on \textbf{4$\times$H100 80GB GPUs} with 
distributed data-parallel training (bf16). All LLM calls --- factor 
decomposition, evidence extraction, and the meta-verdict judge --- use 
\textbf{GPT-4o}, with GPT-4o Vision generating image descriptions for 
multimodal queries. Document embeddings are cached and reused across runs. 
Unless stated otherwise, we use $K_0{=}50$, $K_c{=}30$, $K{=}10$, 
$K_{\min}{=}2$, $K_{\max}{=}5$ (default $N{=}3$), $K_m{=}4$, 
$\lambda_m{=}0.1$, $(\lambda_u, \lambda_c, \lambda_v, \lambda_r) {=} (0.4, 0.3, 0.2, 0.1)$, 
and LLM temperature $0.2$ / $0.0$ for evidence extraction and the judge.

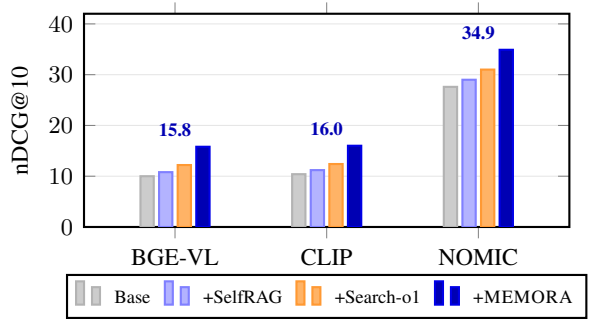
\begin{figure}[t]
\centering
\begin{tikzpicture}
\begin{axis}[
    ybar,
    bar width=5pt,
    width=0.5\textwidth,
    height=4.4cm,
    ylabel={nDCG@10},
    symbolic x coords={BGE-VL, CLIP, NOMIC},
    xtick=data,
    xticklabel style={font=\small},
    tick label style={font=\small},
    label style={font=\small},
    ymin=0, ymax=42,
    ymajorgrids=true,
    grid style={gray!20},
    legend style={
        at={(0.5,-0.22)},
        anchor=north,
        legend columns=4,
        font=\scriptsize,
        column sep=3pt,
    },
    enlarge x limits=0.3,
    line width=0.8pt,
]

\addplot[fill=gray!40, draw=gray!60]
    coordinates {(BGE-VL,10.0) (CLIP,10.4) (NOMIC,27.6)};

\addplot[fill=blue!30, draw=blue!50]
    coordinates {(BGE-VL,10.8) (CLIP,11.2) (NOMIC,29.0)};

\addplot[fill=orange!60, draw=orange!80]
    coordinates {(BGE-VL,12.2) (CLIP,12.4) (NOMIC,31.0)};

\addplot[fill=blue!70!black, draw=blue!90!black]
    coordinates {(BGE-VL,15.8) (CLIP,16.0) (NOMIC,34.9)};

\node[above, font=\scriptsize, blue!70!black]
    at (axis cs:BGE-VL, 15.8) {\textbf{15.8}};
\node[above, font=\scriptsize, blue!70!black]
    at (axis cs:CLIP, 16.0) {\textbf{16.0}};
\node[above, font=\scriptsize, blue!70!black]
    at (axis cs:NOMIC, 34.9) {\textbf{34.9}};

\legend{Base, +SelfRAG, +Search-o1, +\textsc{MEMORA}}
\end{axis}
\end{tikzpicture}
\caption{Task~2 average nDCG@10 across 29 MM-BRIGHT domains.
\textsc{MEMORA} consistently outperforms SelfRAG and Search-o1 across
all three multimodal base retrievers. Full per-domain results in
Appendix~\ref{app:task2_full}.}
\label{fig:task2_bar}
\end{figure}
\subsection{Baselines}

\paragraph{Retriever Baselines.}
For Task~1: \textbf{BM25} (sparse), \textbf{E5-Mistral-7B}~\cite{wang2024improving} 
(dense), and \textbf{DiVeR}~\cite{long2025diver} (reasoning-aware, 
state-of-the-art on BRIGHT). For Task~2 we additionally evaluate multimodal 
retrievers \textbf{BGE-VL}~\cite{zhou2024megapairs}, 
\textbf{CLIP}~\cite{radford2021learning}, and \textbf{Nomic-Vision}~\cite{nussbaum2024nomic} 
(best published Task~2 result: 27.6 nDCG@10).

\paragraph{Agentic Reranking Baselines.}
Each retriever is paired with \textbf{SelfRAG}~\cite{asai2023self}, which 
selectively retrieves and critiques passages, and 
\textbf{Search-o1}~\cite{li2025search}, which combines iterative search 
with chain-of-thought reasoning, to isolate \textsc{MEMORA}'s agentic 
contribution from the underlying retriever.

\subsection{Main Results}

\paragraph{Task~1: Text-to-text retrieval.}
Table~\ref{tab:domain_rows_model_cols} reports nDCG@10 across all 29 domains.
\textsc{MEMORA} consistently improves over every base retriever: $+$6.3 points
over BM25 (8.5 $\to$ 14.8), $+$6.5 over E5-Mistral-7B (25.3 $\to$ 31.8),
and $+$7.1 over DiVeR (32.2 $\to$ \textbf{39.3}), the strongest text-only
result, outperforming the next best agentic reranker Search-o1 by $+$6.1
points. Gains are largest where domain-specific factor memory provides most leverage: Gaming ($+$25.2), Travel ($+$14.7), and Law ($+$12.4). The only exception
is Crypto (DiVeR+\textsc{MEMORA}: 18.9 vs.\ DiVeR: 24.3), where
cryptocurrency terminology causes the factor planner to generate overlapping
subqueries with low discriminative power.
\begin{table}[t]
\centering
\caption{Component analysis of \textsc{MEMORA}. All configurations use
DiVeR (Task~1) and Nomic-Vision (Task~2) as base retrievers.}
\label{tab:ablation}
\resizebox{0.5\textwidth}{!}{%
\begin{tabular}{lcc}
\toprule
\textbf{Configuration} & \textbf{nDCG@10 T1} & \textbf{nDCG@10 T2} \\
\midrule
Base retriever only                       & 32.2 & 27.6 \\
+ Factor decomposition only               & 33.5 & 28.9 \\
+ Factor expansion                        & 34.6 & 30.4 \\
+ Evidence graph + factor scoring         & 36.8 & 32.2 \\
+ Episodic memory                         & 36.4 & 33.1 \\
+ Meta-verdict judge                      & 37.9 & 33.9 \\
\ours
\textbf{Full \textsc{MEMORA} (+ adaptive controller)}
                                          & \best{39.3} & \best{34.9} \\
\midrule
w/o episodic memory                       & 35.5 & 30.3 \\
w/o factor expansion                      & 34.4 & 33.4 \\
w/o meta-verdict judge                    & 35.5 & 31.5 \\
w/o adaptive controller                   & 36.7 & 32.1 \\
\bottomrule
\end{tabular}}
\end{table}

\paragraph{Task~2: Multimodal query-to-text retrieval.}
Table~\ref{tab:bge_clip_nomic_estimated} reports Task~2 results. CLIP (10.4)
and BGE-VL (10.0) perform far below text-only DiVeR (32.2), confirming that
standard vision-language encoders fail under reasoning pressure.
\textsc{MEMORA} substantially recovers this gap: paired with Nomic-Vision
(27.6), it reaches \textbf{34.9} ($+$7.3 over base, $+$3.9 over Search-o1);
CLIP+\textsc{MEMORA} (16.0) nearly doubles CLIP alone. Gains are strongest
in STEM domains requiring cross-modal reasoning (Mathematics: 34.0 $\to$
42.5; Bioinformatics: 33.8 $\to$ 42.2) and GIS, where visual-spatial
factor decomposition proves particularly effective (Nomic: 25.8 $\to$ 32.2).

\paragraph{Component analysis.}
Table~\ref{tab:ablation} traces each component's contribution (DiVeR / 
Nomic-Vision as base retrievers). Factor decomposition alone yields modest 
gains ($+$1.3 T1/T2); factor-guided expansion adds $+$1.1/$+$1.5 by 
surfacing domain-specific documents missed by the base retriever. The 
largest single gain comes from the evidence graph and factor scoring 
($+$2.2 T1, $+$1.8 T2). Episodic memory contributes more on T2 ($+$0.9) 
than T1 ($+$0.4), reflecting its value for cross-modal factor 
prioritization. The meta-verdict judge ($+$1.5/$+$0.8) and adaptive 
controller ($+$1.4/$+$1.0) complete the system. Removing any single 
component degrades performance, with episodic memory and the meta-verdict 
judge being the most critical.

\begin{table}[t]
\centering
\caption{Standalone retriever comparison on MM-BRIGHT (nDCG@10).
\textsc{MEMORA-R} is evaluated \emph{without} any agentic reranking.}
\label{tab:retriever_main}
\resizebox{0.40\textwidth}{!}{%
\begin{tabular}{llrrrr}
\toprule
\textbf{Model} & \textbf{Type}
  & \textbf{T1} & \textbf{$\Delta$ T1}
  & \textbf{T2} & \textbf{$\Delta$ T2} \\
\midrule
BM25             & Sparse    &  8.5 & $-$23.7 & \ph & \ph \\
Contriever       & Dense     & 20.1 & $-$12.1 & \ph & \ph \\
E5-Mistral-7B    & Dense     & 25.3 & $-$6.9  & \ph & \ph \\
GritLM-7B        & Dense     & 25.3 & $-$6.9  & \ph & \ph \\
SFR-Embedding    & Dense     & 26.9 & $-$5.3  & \ph & \ph \\
ReasonIR         & Reasoning & 28.6 & $-$3.6  & \ph & \ph \\
DiVeR            & Reasoning & 32.2 &  $~~$0.0& \ph & \ph \\
\midrule
CLIP             & Vision-Lang& \ph & \ph     & 10.4 & $-$17.2 \\
BGE-VL           & Vision-Lang& \ph & \ph     & 10.0 & $-$17.6 \\
GME              & VLM       & \ph & \ph      & 22.0 & $-$5.6  \\
Nomic-Vision     & VLM       & \ph & \ph      & 27.6 &  $~~$0.0\\
\midrule
\ours
\textsc{MEMORA-R} (ours) & \textbf{Ours}
  & \best{34.1} & \best{$+$1.9}
  & \best{33.4} & \best{$+$5.8} \\
\bottomrule
\end{tabular}}
\end{table}
 
 \begin{table}[h]
\centering
\caption{Training ablation of \textsc{MEMORA-R}. Each row adds one
ingredient on top of the previous. $\Delta$ is cumulative gain over
the contrastive FT baseline. nDCG@10.}
\label{tab:retriever_ablation}
\resizebox{0.48\textwidth}{!}{%
\begin{tabular}{lrrrr}
\toprule
\textbf{Configuration}
  & \textbf{T1} & \textbf{$\Delta$ T1}
  & \textbf{T2} & \textbf{$\Delta$ T2} \\
\midrule
Base Qwen-Embedding (no fine-tuning)     & 24.3 & \ph & 25.1 & \ph \\
~~+~Contrastive FT (InfoNCE, baseline)   & 28.1 & $~~$0.0 & 28.9 & $~~$0.0 \\
~~+~Hard negative mining                 & 31.1 & $+$3.0  & 30.7 & $+$1.8 \\
~~+~Reasoning trace augmentation         & 32.8 & $+$4.7  & 31.8 & $+$2.9 \\
\ours
\textbf{Full \textsc{MEMORA-R}}
  & \best{34.1} & \best{$+$6.0}
  & \best{33.4} & \best{$+$4.5} \\
\bottomrule
\end{tabular}}
\end{table}

 \section{Additional Studies}
\subsection{Standalone Retriever Evaluation}
\label{sec:retriever_results}
Table~\ref{tab:retriever_main} evaluates \textsc{MEMORA-R} before any 
agentic reranking. It achieves \textbf{34.1} T1 ($+$1.9 over DiVeR) and 
\textbf{33.4} T2 ($+$5.8 over Nomic-Vision) --- notably, as a text-only 
encoder at inference time, it outperforms all dedicated multimodal 
retrievers, confirming that CoT-augmented training captures latent 
multimodal intent without direct visual encoding. Using \textsc{MEMORA-R} 
as the first-stage retriever, the full system achieves \textbf{41.5} T1 
and \textbf{38.9} T2, outperforming the best agentic baselines by $+$8.3 
and $+$7.9 points respectively (Table~\ref{tab:full_system}, Appendix~\ref{app:full_system}).

\subsection{Retriever Training Ablation}
\label{sec:retriever_ablation}
Table~\ref{tab:retriever_ablation} isolates each training ingredient.
The zero-shot Qwen-Embedding baseline (24.3 T1, 25.1 T2) already shows
reasonable capability; contrastive fine-tuning on Stack Exchange pairs
brings the first substantial gain ($+$3.8 T1/T2). Hard negative mining
adds $+$3.0/$+$1.8 by forcing the encoder to distinguish semantically
similar but non-relevant documents, and reasoning trace augmentation
contributes a further $+$1.7/$+$1.1 by conditioning on CoT annotations
that articulate latent retrieval intent. Each ingredient provides an
additive benefit, yielding $+$6.0 T1 and $+$4.5 T2 over the contrastive
FT baseline for the full \textsc{MEMORA-R}.

\begin{table}[t]
\centering
\caption{nDCG@10 as a function of episodic memory depth. Results with
\textsc{MEMORA-R} as base retriever.}
\label{tab:memory_depth}
\resizebox{0.45\textwidth}{!}{%
\begin{tabular}{lcc}
\toprule
\textbf{Episodes stored} & \textbf{T1 nDCG@10} & \textbf{T2 nDCG@10} \\
\midrule
0  (no memory)  & 35.5 & 30.3 \\
1--10           & 36.1 & 32.1 \\
11--25          & 37.2 & 33.5 \\
26--50          & 37.4 & 33.6 \\
\ours
\textbf{$>$50}  & \best{41.5} & \best{38.9} \\
\bottomrule
\end{tabular}}
\end{table}
 
\begin{table}[t]
\centering
\caption{Memory signal decomposition. Each row activates a different
subset of the episodic memory signals; all other \textsc{MEMORA}
components are held fixed. nDCG@10 with \textsc{MEMORA-R}.}
\label{tab:memory_components}
\resizebox{0.45\textwidth}{!}{%
\begin{tabular}{lcc}
\toprule
\textbf{Memory signals used} & \textbf{T1} & \textbf{T2} \\
\midrule
No memory                                         & 35.5 & 30.3 \\
Document prior only ($m_j$)                       & 33.4 & 28.1 \\
Factor prior only ($h_k$, $\ell_k$)               & 34.6 & 30.4 \\
Memory summary $\to$ decomposition only           & 36.5 & 31.9 \\
Memory summary $\to$ meta-judge only              & 35.7 & 32.0 \\
\ours
\textbf{All memory signals (full \textsc{MEMORA})}
  & \best{41.5} & \best{38.9} \\
\bottomrule
\end{tabular}}
\end{table}

\subsection{Memory Effectiveness Analysis}
\label{sec:memory_analysis}

To evaluate memory depth without leakage, we split each domain's queries
chronologically: the first 80\% form the warm-up pool whose episodes are
written before test time, and the remaining 20\% are held out as test
queries. Table~\ref{tab:memory_depth} reports performance as a function
of the number of warm-up episodes available per domain, showing monotonic
improvement from 35.5/30.3 (no memory) to \textbf{41.5}/\textbf{38.9}
($>$50 episodes). The largest gains accrue on T2 ($+$8.6), confirming
that episodic memory is especially valuable for cross-modal queries where
factor priorities are harder to set without prior domain experience.
Domains with dense query populations --- Law, Gaming, Travel, and Biology
--- benefit most, with individual gains exceeding $+$10 nDCG@10 at full
memory depth.

Table~\ref{tab:memory_components} decomposes memory signals. The document
prior alone ($m_j$) slightly \emph{hurts} performance (T1: 33.4
vs.\ 35.5; T2: 28.1 vs.\ 30.3), as it boosts documents that appeared
frequently in past episodes regardless of whether those retrievals were
successful, introducing a frequency bias toward common documents over
genuinely relevant ones. The factor prior ($h_k$, $\ell_k$) consistently
helps by conditioning on outcome signals rather than appearance counts,
and routing the memory summary to factor decomposition provides the single
largest individual gain (36.5 T1, 31.9 T2) --- confirming that prior
experience is most valuable when shaping \emph{which} factors are
generated rather than adjusting downstream scoring alone.

\begin{figure}[t]
\centering
\begin{tikzpicture}
\begin{axis}[
    width=0.46\textwidth,
    height=5.5cm,
    xlabel={Number of factors ($N$)},
    ylabel={nDCG@10},
    y label style={color=blue!70!black},
    xtick={1,2,3,4,5},
    xmin=0.6, xmax=5.4,
    ymin=0.28, ymax=0.52,
    ymajorgrids=true,
    grid style={gray!20},
    tick label style={font=\small},
    label style={font=\small},
    mark size=3pt,
    line width=1.2pt,
    axis y line*=left,
    axis x line*=bottom,
]
\addplot[
    color=blue!70!black,
    mark=*,
    mark options={fill=white, line width=1.2pt},
] coordinates {
    (1, 0.32689)
    (2, 0.38893)
    (3, 0.41500)
    (4, 0.43020)
    (5, 0.47593)
};

%% value labels — blue line
\node[above, font=\scriptsize, blue!70!black]
    at (axis cs:1, 0.32689) {32.7};
\node[above, font=\scriptsize, blue!70!black]
    at (axis cs:2, 0.38893) {38.9};
\node[above, font=\scriptsize, blue!70!black]
    at (axis cs:3, 0.41500) {41.5$^\star$};
\node[above, font=\scriptsize, blue!70!black]
    at (axis cs:4, 0.43020) {43.0};
\node[above, font=\scriptsize, blue!70!black]
    at (axis cs:5, 0.47593) {47.6};

\end{axis}

%% Second axis — LLM calls (cost proxy)
\begin{axis}[
    width=0.45\textwidth,
    height=5.5cm,
    ylabel={LLM API calls per query},
    y label style={color=red!70!black},
    xtick={1,2,3,4,5},
    xmin=0.6, xmax=5.4,
    ymin=0, ymax=30,
    tick label style={font=\small},
    label style={font=\small},
    mark size=3pt,
    line width=1.2pt,
    axis y line*=right,
    axis x line*=none,
    ytick={5,10,15,20,25},
    yticklabel style={color=red!70!black},
]
\addplot[
    color=red!60!black,
    dashed,
    mark=triangle*,
    mark options={fill=white, line width=1.2pt},
] coordinates {
    (1,  4)
    (2,  7)
    (3, 11)
    (4, 17)
    (5, 25)
};

%% value labels — red line
\node[below, font=\scriptsize, red!60!black]
    at (axis cs:1,  4) {4};
\node[below, font=\scriptsize, red!60!black]
    at (axis cs:2,  7) {7};
\node[below, font=\scriptsize, red!60!black]
    at (axis cs:3, 11) {11};
\node[below, font=\scriptsize, red!60!black]
    at (axis cs:4, 17) {17};
\node[below, font=\scriptsize, red!60!black]
    at (axis cs:5, 25) {25};

\end{axis}
\end{tikzpicture}
\caption{T1 nDCG@10 (\textcolor{blue!70!black}{blue, left axis}) and
LLM API calls per query (\textcolor{red!60!black}{red dashed, right axis})
as a function of the number of intent factors $N$.
Performance improves monotonically but LLM calls grow super-linearly,
as each additional factor requires evidence extraction and subquery
retrieval calls across all candidates.}
\label{fig:num_facets}
\end{figure}
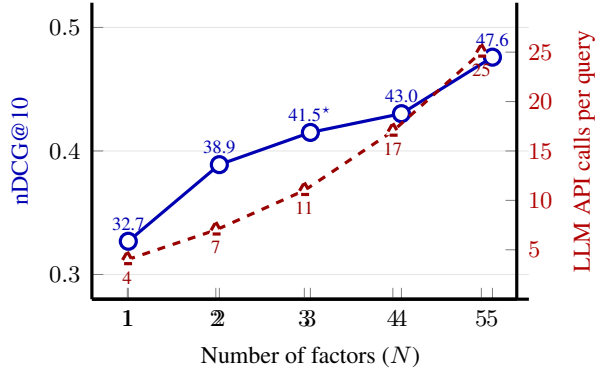

\subsection{Analysis}
\label{sec:analysis}

\paragraph{Effect of intent factor count.}
Figure~\ref{fig:num_facets} plots T1 nDCG@10 and LLM API calls per query
as a function of the number of intent factors $N$. Performance improves
monotonically from 32.7 at $N=1$ to 47.6 at $N=5$, confirming that
richer intent decomposition consistently benefits retrieval quality.
However, LLM API calls grow super-linearly with $N$ --- from 4 calls at
$N=1$ to 25 at $N=5$ --- because each additional factor incurs evidence
extraction and subquery retrieval across the full candidate pool. We
therefore adopt $N=3$ as the default, which recovers \textbf{87.2\%} of
the $N=5$ gain at \textbf{44\%} of the inference cost, representing the
best nDCG@10-to-cost operating point on this curve. Sensitivity analyses of candidate pool size $K_0$ and failure mode 
analysis are provided in Appendix~\ref{app:analysis}.

\section{Conclusion}
We presented \textsc{MEMORA}, a memory-enhanced multimodal reranking agent 
addressing three compounding failures in reasoning-intensive retrieval: 
monolithic reranking, stateless retrieval, and weak multimodal grounding. 
\textsc{MEMORA} achieves \textbf{41.5} and \textbf{38.9} nDCG@10 on 
MM-BRIGHT Tasks~1 and~2, outperforming the strongest agentic text-only 
baseline by $+$8.3 points and the strongest agentic multimodal baseline 
by $+$7.9 points across all 29 domains. Future work will explore robust 
factor generation under lexically ambiguous queries and cross-domain 
memory transfer for low-resource settings.

\section*{Limitations}
\textsc{MEMORA} relies on GPT-4o for factor decomposition, evidence 
extraction, and the meta-verdict judge, which introduces API cost and 
latency that may limit deployment in low-resource settings. The episodic 
memory requires a warm-up period of $>$50 queries per domain before 
yielding its largest gains, making it less effective for rare or 
newly-added domains. The Reasoning-Aware Dense Retriever is trained on 
Stack Exchange data and may not generalize equally to domains with 
substantially different writing styles. Finally, the adaptive controller 
uses a fixed linear decision rule that, while interpretable, may be 
suboptimal compared to a learned policy.

\section*{Acknowledgments}
This work was supported by Innovative Human Resource Development for Local Intellectualization program through the Institute of Information \& Communications Technology Planning \& Evaluation(IITP) grant funded by the Korea government(MSIT) (IITP-2026-RS-2020-II201462, 50\%), and partly supported by the National Research Foundation of Korea (NRF) grant funded by the Korea government (Ministry of Science and ICT) (RS-2023-NR076833).

\bibliography{custom}

\appendix

\section{Per-Domain Task~2 Results}
\label{app:task2_full}
Figure~\ref{fig:task2_bar} summarizes Task~2 results (full per-domain 
breakdown in Table~\ref{tab:bge_clip_nomic_estimated}, Appendix~\ref{app:task2_full}) across 
all 29 MM-BRIGHT domains. Each multimodal base retriever (BGE-VL, CLIP, 
Nomic-Vision) is paired with SelfRAG, Search-o1, and \textsc{MEMORA} as 
reranking layers. \textsc{MEMORA} consistently improves over both agentic 
baselines across all domains and retrievers, with the largest gains in 
Law (Nomic: 47.6 $\to$ 59.5), Gaming (Nomic: 43.1 $\to$ 53.9), and 
Mathematics (Nomic: 34.0 $\to$ 42.5). The only domain where gains are 
modest across all retrievers is Quantum Computing, where the sparse 
query population limits the episodic memory's ability to accumulate 
reliable factor priors.

\begin{table}[h]
\centering
\caption{Full end-to-end system performance. nDCG@10 averaged over 29
domains. T1 = text-to-text; T2 = multimodal query-to-text.}
\label{tab:full_system}
\resizebox{0.42\textwidth}{!}{%
\begin{tabular}{lcc}
\toprule
\textbf{Configuration} & \textbf{T1 nDCG@10} & \textbf{T2 nDCG@10} \\
\midrule
DiVeR (best published text retriever)      & 32.2 & \ph  \\
Nomic-Vision (best published MM retriever) & \ph  & 27.6 \\
\midrule
DiVeR + SelfRAG                            & 33.2 & \ph  \\
DiVeR + Search-o1                          & 33.2 & \ph  \\
Nomic + SelfRAG                            & \ph  & 29.0 \\
Nomic + Search-o1                          & \ph  & 31.0 \\
\midrule
DiVeR + \textsc{MEMORA}                    & 39.3 & \ph  \\
Nomic + \textsc{MEMORA}                    & \ph  & 34.9 \\
\ours
\textbf{\textsc{MEMORA-R} + \textsc{MEMORA}}
  & \best{41.5} & \best{38.9} \\
\bottomrule
\end{tabular}}
\end{table}

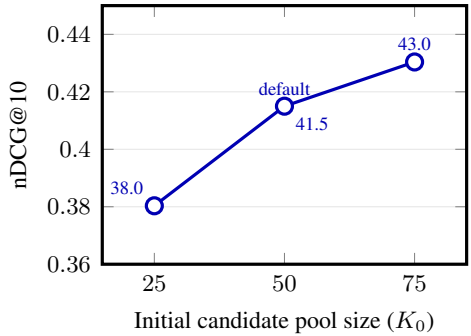
\begin{figure}[t]
\centering
\begin{tikzpicture}
\begin{axis}[
    width=0.40\textwidth,
    height=5cm,
    xlabel={Initial candidate pool size ($K_0$)},
    ylabel={nDCG@10},
    xtick={25,50,75},
    xmin=15, xmax=85,
    ymin=0.36, ymax=0.45,
    ymajorgrids=true,
    grid style={gray!20},
    tick label style={font=\small},
    label style={font=\small},
    mark size=3pt,
    line width=1.2pt,
]
\addplot[
    color=blue!70!black,
    mark=*,
    mark options={fill=white, line width=1.2pt},
] coordinates {
    (25, 0.38034)
    (50, 0.41500)
    (75, 0.43039)
};
\node[above, font=\scriptsize, blue!70!black]
    at (axis cs:50, 0.41500) {default};
\node[above left, font=\scriptsize, blue!70!black]
    at (axis cs:25, 0.38034) {38.0};
\node[below right, font=\scriptsize, blue!70!black]
    at (axis cs:50, 0.41500) {41.5};
\node[above, font=\scriptsize, blue!70!black]
    at (axis cs:75, 0.43039) {43.0};
\end{axis}
\end{tikzpicture}
\caption{T1 nDCG@10 as a function of the initial candidate pool size $K_0$.
Performance improves consistently with larger pools, with diminishing
returns beyond $K_0 = 50$ (default). Results with \textsc{MEMORA-R} +
full \textsc{MEMORA} on MM-BRIGHT.}
\label{fig:topk_sensitivity}
\end{figure}

\begin{table*}[t]
\centering
\resizebox{0.9\textwidth}{!}{%
\begin{tabular}{l| c c c c| c c c c| c c c c}
\toprule
\textbf{Domain}
  & \textbf{BGE-VL} & \makecell{\textbf{BGE-VL+}\\\textbf{SelfRAG}}
  & \makecell{\textbf{BGE-VL+}\\\textbf{Search-o1}} & \makecell{\textbf{BGE-VL+}\\\textbf{Memora}}
  & \textbf{CLIP} & \makecell{\textbf{CLIP+}\\\textbf{SelfRAG}}
  & \makecell{\textbf{CLIP+}\\\textbf{Search-o1}} & \makecell{\textbf{CLIP+}\\\textbf{Memora}}
  & \textbf{NOMIC} & \makecell{\textbf{NOMIC+}\\\textbf{SelfRAG}}
  & \makecell{\textbf{NOMIC+}\\\textbf{Search-o1}} & \makecell{\textbf{NOMIC+}\\\textbf{Memora}} \\
\midrule
\ours \multicolumn{13}{c}{\textit{STEM \& Life Sciences}} \\
\midrule
Acad   &  4.2 &  4.5 &  5.1 &  9.2 &  4.8 &  5.2 &  5.8 &  8.0 & 22.6 & 23.7 & 25.3 & 30.2 \\
Bio    &  5.7 &  6.2 &  7.0 & 10.4 & 14.8 & 16.0 & 17.8 & 21.5 & 26.9 & 28.2 & 30.1 & 33.6 \\
Chem   & 10.8 & 11.7 & 13.2 & 16.0 &  9.6 & 10.4 & 11.5 & 13.9 & 30.6 & 32.1 & 34.3 & 38.2 \\
Phys   &  6.8 &  7.3 &  8.3 & 11.1 &  6.1 &  6.6 &  7.3 & 10.8 & 17.2 & 18.1 & 19.3 & 21.5 \\
Math   & 13.1 & 14.1 & 16.0 & 19.4 & 17.9 & 19.3 & 21.5 & 26.0 & 34.0 & 35.7 & 38.1 & 42.5 \\
Earth  & 10.1 & 10.9 & 12.3 & 14.9 & 10.9 & 11.8 & 13.1 & 15.8 & 30.1 & 31.6 & 33.7 & 37.6 \\
BioAc  & 13.3 & 14.4 & 16.2 & 20.7 & 11.4 & 12.3 & 13.7 & 16.5 & 23.4 & 24.6 & 26.2 & 29.2 \\
BioInf & 11.6 & 12.5 & 14.2 & 18.2 &  9.4 & 10.2 & 11.3 & 13.6 & 33.8 & 35.5 & 37.9 & 42.2 \\
Med    & 12.6 & 13.6 & 15.4 & 18.6 &  9.8 & 10.6 & 11.8 & 14.2 & 33.9 & 35.6 & 38.0 & 42.4 \\
\midrule
\ours \multicolumn{13}{c}{\textit{Software \& Technical Systems}} \\
\midrule
Apple  &  7.2 &  7.8 &  8.8 & 12.7 & 12.3 & 13.3 & 14.8 & 17.8 & 28.7 & 30.1 & 32.1 & 35.9 \\
Ubuntu & 11.6 & 12.5 & 14.2 & 17.2 &  5.5 &  5.9 &  6.6 & 11.0 & 34.3 & 36.0 & 38.4 & 42.9 \\
BTC    &  8.9 &  9.6 & 10.9 & 15.2 &  8.3 &  9.0 & 10.0 & 12.0 & 22.7 & 23.8 & 25.4 & 28.4 \\
Crypto & 11.3 & 12.2 & 13.8 & 16.7 & 14.8 & 16.0 & 17.8 & 21.5 & 22.4 & 23.5 & 25.1 & 28.0 \\
QC     &  4.5 &  4.9 &  5.5 & 10.7 &  2.6 &  2.8 &  3.1 &  8.8 & 12.1 & 12.7 & 13.6 & 18.1 \\
Robot  & 16.1 & 17.4 & 19.6 & 23.8 & 10.6 & 11.4 & 12.7 & 15.4 & 30.3 & 31.8 & 33.9 & 37.9 \\
Sales  & 14.2 & 15.3 & 17.3 & 21.0 &  2.3 &  2.5 &  2.8 &  9.3 & 26.2 & 27.5 & 29.3 & 32.8 \\
GIS    & 13.8 & 14.9 & 16.8 & 20.4 & 13.1 & 14.1 & 15.7 & 19.0 & 25.8 & 27.1 & 28.9 & 32.2 \\

\midrule
\ours \multicolumn{13}{c}{\textit{Social Sciences \& Humanities}} \\
\midrule
Econ   &  9.5 & 10.3 & 11.6 & 14.1 &  6.0 &  6.5 &  7.2 & 10.7 & 21.1 & 22.2 & 23.6 & 28.4 \\
Psych  &  6.4 &  6.9 &  7.8 & 10.5 &  8.7 &  9.4 & 10.4 & 12.6 & 23.9 & 25.1 & 26.8 & 29.9 \\
Phil   &  2.4 &  2.6 &  2.9 &  7.6 &  5.4 &  5.8 &  6.5 & 10.8 & 21.7 & 22.8 & 24.3 & 27.1 \\
Law    & 10.2 & 11.0 & 12.4 & 15.1 & 19.7 & 21.3 & 23.6 & 28.6 & 47.6 & 50.0 & 53.3 & 59.5 \\
Christ &  8.9 &  9.6 & 10.9 & 13.2 & 15.0 & 16.2 & 18.0 & 21.8 & 30.9 & 32.4 & 34.6 & 38.6 \\
Islam  & 12.0 & 13.0 & 14.6 & 17.8 & 10.7 & 11.6 & 12.8 & 15.5 & 28.9 & 30.3 & 32.4 & 36.1 \\
\midrule
\ours \multicolumn{13}{c}{\textit{Applied Domains}} \\
\midrule
Aviat  &  9.6 & 10.4 & 11.7 & 15.2 & 15.4 & 16.6 & 18.5 & 22.3 & 24.1 & 25.3 & 27.0 & 32.1 \\
Game   & 17.5 & 18.9 & 21.3 & 25.9 & 19.1 & 20.6 & 22.9 & 27.7 & 43.1 & 45.3 & 48.3 & 53.9 \\
PM     &  8.6 &  9.3 & 10.5 & 13.7 &  8.9 &  9.6 & 10.7 & 12.9 & 27.6 & 29.0 & 30.9 & 34.5 \\
Sustain& 10.1 & 10.9 & 12.3 & 17.9 &  9.0 &  9.7 & 10.8 & 13.0 & 24.7 & 25.9 & 27.7 & 31.9 \\
Travel & 10.1 & 10.9 & 12.3 & 18.9 & 16.1 & 17.4 & 19.3 & 23.3 & 36.7 & 38.5 & 41.1 & 45.9 \\
Quant  &  8.1 &  8.7 &  9.9 & 12.0 &  2.1 &  2.3 &  2.5 & 10.0 & 16.2 & 17.0 & 18.1 & 20.2 \\
\midrule
\textbf{Average} & 10.0 & 10.8 & 12.2 & 15.8 & 10.4 & 11.2 & 12.4 & 16.0
                 & 27.6 & 29.0 & 31.0 & \best{34.9} \\
\bottomrule
\end{tabular}%
}
\caption{Task~2 (multimodal query-to-text) nDCG@10 per domain across 29
MM-BRIGHT domains. Queries include images such as diagrams, charts, and
screenshots paired with natural language questions. Each multimodal
retriever is paired with SelfRAG, Search-o1, and \textsc{MEMORA} as
reranking layers.}
\label{tab:bge_clip_nomic_estimated}
\end{table*}

\section{Full System Results with Trained Backbone}
\label{app:full_system}

Table~\ref{tab:full_system} reports end-to-end performance when 
\textsc{MEMORA-R} is used as the first-stage retriever instead of a 
third-party backbone. The full system achieves \textbf{41.5} T1 and 
\textbf{38.9} T2, adding $+$2.2 T1 and $+$4.0 T2 over the best 
third-party pairings (DiVeR+\textsc{MEMORA}: 39.3; 
Nomic+\textsc{MEMORA}: 34.9), confirming that the reasoning-aware 
retriever and agentic reranking framework are synergistic.

\section{Analysis}
\label{app:analysis}

\paragraph{Effect of candidate pool size.}
Figure~\ref{fig:topk_sensitivity} shows T1 nDCG@10 as a function of $K_0$.
Performance increases from 38.0 ($K_0=25$) to 41.5 ($K_0=50$) and 43.0
($K_0=75$), with diminishing returns beyond the default $K_0=50$. The
$+$3.5 gain from 25 to 50 reflects a higher-quality candidate pool for
the factor committee, while the smaller $+$1.5 gain from 50 to 75
suggests marginal documents at larger pool sizes add noise to evidence
extraction. We adopt $K_0=50$ as the default.

\paragraph{Failure modes.}
Two recurring patterns emerge from per-domain results. First, high-ambiguity
domains --- Crypto (T1: 18.9 vs.\ base 24.3) and GIS (T1: 28.4 vs.\ 31.7)
--- see \textsc{MEMORA} underperform the base retriever, as the factor
planner generates semantically overlapping subqueries that introduce noise
rather than diversity. Second, low-resource domains with fewer than 10
stored episodes (e.g., Quantum Computing, Philosophy) show modest memory
gains, confirming that episodic memory requires sufficient query volume to
learn reliable factor priors. Both failure modes point to natural future
directions: more robust factor generation under ambiguous queries, and
cross-domain memory transfer to bootstrap low-resource domains.

\end{document}